\documentclass[twocolumn,twoside,russian,english]{article}

\usepackage{graphicx}
\usepackage{hyperref}
\begin{document}

\thispagestyle{plain}
\markboth{\rm \uppercase{Are black holes totally black?}}{\rm
\uppercase{Grib, Pavlov}}

\twocolumn[
\begin{center}
{\LARGE \bf Are Black Holes Totally Black?}
\vspace{12pt}

{\large \bf {A. A. Grib${}^{1,2,3,*}$, \ Yu. V. Pavlov${}^{1,2,4,**}$}}
\vspace{12pt}

{${}^{1}$\it A.\,Friedmann Laboratory for Theoretical Physics,
St.\,Petersburg, Russia;}
\vspace{4pt}

{\it ${}^{2}$Copernicus Center for Interdisciplinary Studies,
Krak\'{o}w, Poland}
\vspace{4pt}

{\it ${}^{3}$Theoretical Physics and Astronomy Department,
The Herzen University,\\
Moika 48, St. Petersburg 191186, Russia;}
\vspace{4pt}

{\it ${}^{4}$Institute of Problems in Mechanical Engineering,
Russian Acad. Sci.,\\
Bol'shoy pr. 61, St. Petersburg 199178, Russia;}
\end{center}
\vspace{5pt}
 {\bf Abstract.}
     Geodesic completeness needs existence near the horizon of the black hole
of ``white hole'' geodesics coming from the region inside of the horizon.
    Here we give the classification of all such geodesics with the energies
$E/m \le 1$ for the Schwarzschild and Kerr's black hole.
    The collisions of particles moving along the ``white hole'' geodesics with
those moving along ``black hole'' geodesics are considered.
    Formulas for the increase of the energy of collision in the centre of mass
frame are obtained and the possibility of observation of high energy
particles arriving from the black hole to the Earth is discussed.

\vspace{11pt}
\vspace{17pt}
]

%%%%******************************************************************
{\centering  \section{Introduction}}

\footnotetext[1]{andrei\_grib@mail.ru}
\footnotetext[7]{yuri.pavlov@mail.ru}

    It is well known~\cite{LL_II} that in order to have the geodesic completeness
in the vicinity of the horizon of the black holes one cannot consider only
those geodesics which go inside the horizon.
    Imagine the situation when somebody is at rest in space relative to the black
hole not far from the horizon.
    Surely he (she) must have strong acceleration in order not to fall to
the black hole.
    Then he (she) throws a stone.
    The stone is falling along the geodesic towards the black hole.
    This geodesic is characterized by the energy
$E= \sqrt{g_{00}} m c^2 $, so that $E/m<1$.
    All geodesics coming to the black hole from space infinity are
such~\cite{Chandrasekhar} that $E/m>1$.
    Geodesics must originate either in infinity or in singularity.
    So what is the origin of geodesics with $E/m<1$?
    The answer is well known~\cite{LL_II}.
    These geodesics must come from the region inside of the horizon.
    They arise either in white hole singularity or in the infinity of the other
universe in full Schwarzschild and Kerr's solutions as combination of the black
and white holes.
    So we want to stress the usually neglected fact that ``eternal'' black
holes with geodesics going only inside the horizon cannot exist!

    Recently~\cite{GribPavlovVert} we showed that for the case of rotating
black holes described by Kerr's metric geodesics characterized by the negative
energy (Penrose geodesics) leading to the Penrose effect originate inside
the horizon and are ``white hole'' geodesics.

    In this paper we give the classification of all such geodesics with
the positive and negative energies.
    In Sec.~\hyperref[Sec2]{2} we give formulas for the energies
characterizing these geodesics for the Schwarzschild black hole and
formulas for the maximal distances from the black hole after which
they turn back to the black hole.
    In Sec.~\hyperref[Sec3]{3} the same analysis is made for the Kerr's case.

    If one considers some massive or massless particles moving along
``white hole'' geodesics then the collision of these particles with ordinary
particles moving along the ``black hole'' geodesics leads to the effect of
``supercollider'' in the vicinity of the horizon.
    Formulas for the energy of the colliding particles in the centre of mass
frame and its dependence on the distance from the horizon are
obtained in Sec.~\hyperref[Sec4]{4}.
    This effect is similar to the well known Banados-Silk-West (BSW)
effect~\cite{BanadosSilkWest09} studied by us previously for multiple
collisions close to the horizon of the rotating black
holes~\cite{GribPavlov2010}---\cite{GribPavlovPiattella2012}
and at any point of ergosphere depending on angular momentum
in~\cite{GribPavlov2012,GribPavlov2013}.
    In \hyperref[SecConcl]{Conclusion} the problem of cosmic censorship and
the consequences for the case of the collapse of stars are shortly discussed.

    We use the units with $G=c=1$.

\vspace{1em}
%%%% *****************************************************************
{\centering  \section{Nonrotating black hole}
\label{Sec2}}

    The Schwarzschild metric for nonrotating black hole has the form
    \begin{eqnarray}
d s^2 &=& \biggl( 1 - \frac{2M}{r} \biggr) d t^2 -
\frac{ d r^2 }{\displaystyle 1 - \frac{2M}{r} }
\nonumber \\
&& -\, r^2 \! \left( \sin^2\! \theta \, d \varphi^2 + d \theta^2 \right),
\label{g1}
\end{eqnarray}
    where $M$ is the mass of the black hole.
    The event horizon corresponds to $r=r_H \equiv 2 M$.

    The equations for geodesics in metric~(\ref{g1}) can be written
for $\theta =0$ as
    \begin{equation} \label{geodSch1}
\frac{d t}{d \lambda} = \frac{x}{x-2}\, E,
\end{equation}
    \begin{equation}
\frac{d \varphi}{d \lambda} = \frac{J}{r^2},
\label{geodSch2}
\end{equation}
    \begin{equation} \label{geodSch3}
\left( \frac{d r}{d \lambda} \right)^2 =
E^2 + \frac{2 M - r}{r^3} J^2 + \frac{2 M - r}{r} m^2,
\end{equation}
    where $E$ is the energy of the moving particle,
$J$ --- the conserved projection of the particle angular momentum on the
axis orthogonal to the plane of movement,
$m$ is the mass of the test particle.
    For particle with nonzero rest mass $\lambda = \tau/m$,
where $\tau$ is the proper time of the massive particle.

    Define the effective potential by the formula
    \begin{equation} \label{Leff}
V_{\rm eff} = -\frac{1}{2} \left[
E^2 + \frac{2 M - r}{r^3} J^2 + \frac{2 M - r}{r} m^2 \right].
\end{equation}
    Then
    \begin{equation} \label{LeffUR}
\frac{1}{2} \left( \frac{d r}{d \lambda} \right)^{\!2} + V_{\rm eff}=0, \ \ \ \
\frac{d^2 r}{d \lambda^2} = - \frac{d V_{\rm eff}}{d r}.
\end{equation}
    The permitted region of particle movement is defined by the condition
    \begin{equation} \label{VG0}
V_{\rm eff} \le 0
\end{equation}
    and by the condition of movement ``forward in time''
    \begin{equation} \label{Vtl0}
d t / d \lambda > 0 .
\end{equation}
    The last condition leads to the positivity of the energy $E>0$
for the region outside of the horizon~\cite{GribPavlov2010NE}.

    For particles with nonzero rest mass and the special
energy $\varepsilon = E /m $ smaller than one
the permitted region of movement occurs to be limited by
    \begin{equation} \label{xRazr}
0 < \varepsilon < 1 \ \ \Rightarrow \ \ r \le \frac{2 M}{1 - \varepsilon^2}
\end{equation}
    as it follows from~(\ref{Leff}) and~(\ref{VG0}).

    Let us prove that trajectories of particles with the energy
$\varepsilon < 2 \sqrt{2}/3 $
outside of the horizon of the Schwarzschild black hole originate and
terminate at $r=r_H$.
    For this it is sufficient to prove that
    \begin{equation} \label{ff}
0 < \varepsilon < \frac{2 \sqrt{2} }{3},  \ r > r_H, \
V_{\rm eff}(r) = 0 \ \ \Rightarrow \ \ \frac{d V_{\rm eff}}{d r} > 0 .
\end{equation}
    This means that at the upward point of the trajectory
(with maximal value of~$r$) the radial acceleration $d^2 r / d \lambda^2 $
is negative~(\ref{LeffUR})
and the downward boundary is absent up to the horizon.
    The boundary value of the specific energy in~(\ref{ff})
corresponds to the minimal value of the specific energy on the orbits
with constant value of the radial coordinate due to the fact
that the condition existence of such orbits is
    \begin{equation} \label{LeffCucl}
V_{\rm eff}=0, \ \ \ \ \frac{d V_{\rm eff}}{d r} =0\,.
\end{equation}
    In fact, for the energy lower than the minimal energy on the orbit
with constant value of~$r$ the particle at the upward point of the
trajectory outside the event horizon of the black hole must have
negative radial acceleration, i.e. $d V_{\rm eff} / d r >0$.

    These orbits for the Schwarzschild black hole are circular.
    For this case from the system of the equation~(\ref{LeffCucl})
one can easily obtain for circular orbits the conditions
    \begin{equation} \label{ElKrug}
\varepsilon^2 = \frac{(r-2M)^2}{r (r-3M)} , \ \ \ \
l^2 = \frac{r^2}{M(r-3M)},
\end{equation}
    where $l= J/(mM)$ is the dimensionless specific projection of the angular
momentum of the particle.
    From~(\ref{ElKrug}) one obtains the minimal value of the energy of
particles on circular orbits $\varepsilon_0 = 2 \sqrt{2}/3$,
which proves our conjecture~(\ref{ff}).
    The corresponding circular orbit has $r= 6 M$
and is stable orbit with the minimal radius~(\cite{LL_II}, \S~102).

    Note that the possible region of movement of particles with
the specific energy $\varepsilon < 2 \sqrt{2} / 3 $
due to the formula~(\ref{xRazr}) leads to $r < 18M = 9 r_H$.
     The maximal values of~$r$ are obtained by particles with
zero value of the angular momentum.

\vspace{2em}
%%%% *****************************************************************
{\centering  \section{The limiting energies for ``white-black'' hole
geodesics in Kerr's metric}
\label{Sec3}}

    Kerr's metric~\cite{Kerr63} of the rotating black hole in
Boyer-Lindquist~\cite{BoyerLindquist67} coordinates has the form

    \begin{eqnarray}
d s^2 &=& d t^2 -
\frac{2 M r}{\rho^2} \, ( d t - a \sin^2 \! \theta\, d \varphi )^2
\label{Kerr} \\
&&-\, \rho^2 \left( \frac{d r^2}{\Delta} + d \theta^2 \right)
- (r^2 + a^2) \sin^2 \! \theta\, d \varphi^2,
\nonumber
\end{eqnarray}
    where
    \begin{equation} \label{Delta}
\rho^2 = r^2 + a^2 \cos^2 \! \theta, \ \ \ \ \
\Delta = r^2 - 2 M r + a^2,
\end{equation}
    $M$ is the mass of the black hole, $aM$ --- its angular momentum.
    The rotation axis direction corresponds to $\theta =0$, i.e. $a \ge 0$.
    The event horizon of the Kerr's black hole corresponds to
    \begin{equation}
r = r_H \equiv M + \sqrt{M^2 - a^2} .
\label{Hor}
\end{equation}
    The surface of the static limit is defined by
    \begin{equation}
r = r_1(\theta) \equiv M + \sqrt{M^2 - a^2 \cos^2 \theta} .
\label{Lst}
\end{equation}
    In case $ a \le M $ the region of space-time between the static limit
and event horizon is called ergosphere.

    For geodesics in Kerr's metric~(\ref{Kerr}) one obtains
(see Ref.~\cite{Chandrasekhar}, Sec.~62 or
Ref.~\cite{NovikovFrolov}, Sec.~3.4.1)
    \begin{equation} \label{geodKerr1}
\rho^2 \frac{d t}{d \lambda } = -a \left( a E \sin^2 \! \theta - J \right)
+ \frac{r^2 + a^2}{\Delta}\, P,
\end{equation}
    \begin{equation}
\rho^2 \frac{d \varphi}{d \lambda } =
- \left( a E - \frac{J}{\sin^2 \! \theta} \right) + \frac{a P}{\Delta} ,
\label{geodKerr2}
\end{equation}
    \begin{equation} \label{geodKerr3}
\rho^2 \left( \frac{d r}{d \lambda} \right) = \sigma_r \sqrt{R}, \ \ \ \
\rho^2 \left(  \frac{d \theta}{d \lambda} \right) =\sigma_\theta \sqrt{\Theta},
\end{equation}
    \begin{equation} \label{geodP}
P = \left( r^2 + a^2 \right) E - a J,
\end{equation}
    \begin{equation} \label{geodR}
R = P^2 - \Delta [ m^2 r^2 + (J- a E)^2 + Q],
\end{equation}
    \begin{equation} \label{geodTh}
\Theta = Q - \cos^2 \! \theta \left[ a^2 ( m^2 - E^2) +
\frac{J^2}{\sin^2 \! \theta} \right].
\end{equation}
    Here $E$ is conserved energy (relative to infinity)
of the probe particle,
$J$ is conserved angular momentum projection on the rotation axis
of the black hole,
$m$ is the rest mass of the probe particle, for particles with nonzero
rest mass $\lambda = \tau /m $,
where $\tau$ is the proper time for massive particle,
$Q$ is the Carter's constant.
    The constants $\sigma_r, \sigma_\theta = \pm 1$ define the direction of
particles movement in coordinates $r, \theta$.

%%%%%%%%%%%%%%%%%%%%%%%%%%%%%%%%%%%%%%%%%%%%%%%%%%
    The permitted region for particle movement is defined by conditions
    \begin{equation} \label{ThB0}
R \ge 0, \ \ \ \ \ \Theta \ge 0, \ \ \ \ \
\frac{d t}{d \lambda} \ge 0 .
\end{equation}
    The corresponding permitted values for the energy and the angular
momentum of particles are given in~\cite{GribPavlov2013}.
    Note that  from~(\ref{ThB0}) close to the horizon
for $\theta \ne 0, \pi$ one obtains
    \begin{equation} \label{JgEH}
r \to r_H \ \ \Rightarrow \ \
J \le J_H = \frac{ 2 M r_H E}{a} .
\end{equation}
    So $J_H$ is the maximal value of the angular momentum of the particle with
the energy~$E$ close to the gravitational radius.

    As we did it before to find the limiting value of the energy for
the geodesics originating and terminating at $r=r_H$
let us find the minimal value of the energy for geodesic orbits with
constant value of the radial coordinate~$r$.

    The effective potential~(\ref{LeffUR}) in case of the Kerr's metric is
    \begin{equation} \label{LeffK}
V_{\rm eff} = - \frac{R}{2 \rho^4}.
\end{equation}
    The conditions of the existence of orbits with the constant value of the
radial coordinate~(\ref{LeffCucl}) (the spherical orbits) for particles with
nonzero rest mass can be written in the form
    \begin{equation} \label{NFore}
\varepsilon = \frac{x^3 (x-2) - A^2 q + A F}
{x^2 \sqrt{x^3(x-3) - 2 A^2 q + 2 A F } }
\end{equation}
    \begin{equation} \label{NForl}
l = \frac{(x^2 + A^2)
\left( F - A q \right) - 2 A x^3 }
{x^2 \sqrt{x^3(x-3) - 2 A^2 q + 2 A F } }\,,
\end{equation}
    where
    \begin{equation} \label{xAq}
x = \frac{r}{M} , \ \ \ \ A = \frac{a}{M} , \ \ \ \ q = \frac{Q}{m^2 M^2} ,
\end{equation}
    \begin{equation} \label{xAF}
F = \pm \sqrt{x^5 + A^2 q^2 + q x^3 (3-x)} ,
\end{equation}
    the upper sign corresponds to the direct orbits (i.e., the projection of
orbital angular momentum of a particle on the axis of rotating of the black
hole is positive), the lower sign corresponds to retrograde orbits.
    To obtain the formulas~(\ref{NFore}), (\ref{NForl})
from equations~(\ref{LeffCucl}), (\ref{LeffK})
one must use elementary but considerable algebraic transformations.
    The formulas~(\ref{NFore}), (\ref{NForl}) are generalizations of
the known expressions~\cite{BardeenPressTeukolsky72} for circular
orbits in equatorial plane ($Q=0$):
    \begin{equation} \label{ShapTyu1a}
\varepsilon = \frac{x^{3/2} - 2 \sqrt{x} \pm A }
{\sqrt{x \left( x^2- 3 x \pm 2 A \sqrt{x}\, \right) }}\,,
\end{equation}
    \begin{equation} \label{ShapTyu1b}
l = \pm \frac{x^2 \mp 2 A \sqrt{x} + A^2 }
{\sqrt{x \left( x^2- 3 x \pm 2 A \sqrt{x}\, \right) }}\,.
\end{equation}
    Further we shall consider only the case of movement in the equatorial
plane of the black hole.

\begin{figure}[ht]
\centering
\includegraphics[width=72mm]{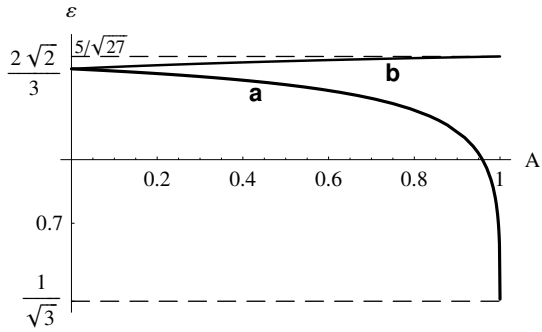}
\caption{The limiting values of energies.
The line~(\textbf{a}) describes the case of particles rotating in the same
direction that of the black hole, the line~(\textbf{b}) --- in the inverse
direction.}
\label{FigGr1}
\end{figure}

    To calculate the minimal value of the specific energy on circular orbits
let us find the extremum of~(\ref{ShapTyu1a}) corresponding to the real
root of the equation
    \begin{equation} \label{Vk}
x^2 - 6 x \pm 8 A \sqrt{x} - 3 A^2 = 0.
\end{equation}
    This root coincides with the value of the minimal radius of the stable
circular orbit, first found in~\cite{BardeenPressTeukolsky72}:
    \begin{equation} \label{ShapTyu4}
x_{\rm ms}^\pm = 3+ Z_2 \mp \sqrt{ (3-Z_1) ( 3+ Z_1 + 2 Z_2) }\,,
\end{equation}
    where
    \begin{eqnarray}
&&Z_1 = 1 + \left( 1\!-\!A^2 \right)^{1/3} \left[ (1\!+\! A)^{1/3} +
(1\!-\!A)^{1/3} \right] , \nonumber \\
&&Z_2 = \sqrt{ 3 A^2 + Z_1^2} .
\label{ShapTyu5}
\end{eqnarray}
    The specific energy of the particle on such a limiting stable orbit is
$\varepsilon = \sqrt{1-  2 /(3 x_{\rm ms})}$.
    The graph for the limiting values of the energy is represented on
Fig.~\ref{FigGr1}.
%%%%%%%%%%%%%%%%%%%%%%%%%%%%%%%%%%%%%%%%%%%%%%%%%
    The minimal value of the energy on the circular orbit takes place
for the plus sign in~(\ref{ShapTyu4}),
i.e. for direct orbits when the particle is rotating in the same direction
as the black hole.

    For the extremal rotating black hole $A=1$ the
calculations~(\ref{ShapTyu4}) give the well known
result~(\cite{LL_II}, \S~104)
    \begin{equation} \label{A1xe}
A=1 \Rightarrow \
x_{\rm ms}^+ =1, \ \varepsilon^+ \!= \frac{1}{\sqrt{3}}, \
x_{\rm ms}^- =9 , \ \varepsilon^- \!= \frac{5}{3 \sqrt{3}} .
\end{equation}

    So in Kerr's metric all geodesics with specific energy
    \begin{equation} \label{AxeM}
\varepsilon < \sqrt{1-  \frac{2}{3 x_{\rm ms}^+} }
\end{equation}
    with $x_{\rm ms}^+$ defined by~(\ref{ShapTyu4}) originate and
terminate at $r=r_H$.
    If the angular momentum of the black hole is growing from $a=0$
to $a=M$ the limiting energy is changing from $2 \sqrt{2}/3$ for $a=0$
to $1 / \sqrt{3} $ for $a=M$
(see the lower line~(\textbf{a}) on Fig.~\ref{FigGr1}).

    If one consider geodesics only for retrograde orbits (negative angular
momentum projection) one can say that all such geodesics with specific energies
$\varepsilon < \sqrt{1-  2 /(3 x_{\rm ms}^-)}$
originate and terminate at $r=r_H$.
    If the angular momentum of the black hole is growing the limitating
energy is growing from $2 \sqrt{2}/3$ ($a=0$) to $5 / (3 \sqrt{3}) $ ($a=M$)
(see the upper line~(\textbf{b}) on Fig.~\ref{FigGr1}).

    Let us find the possible region of particle movement in the equatorial
plane for the given specific energy.
    Due to formula~(11) of our paper~\cite{GribPavlov2012}
one obtains the same limit~(\ref{xRazr}) for the region outside the ergosphere
that exists for the nonrotating black hole.
    It is evident that the upper boundary of the permitted region of
particle movement with positive energy is always located out of the ergosphere.
    Note that the maximal value~$r_{\rm max}$ for given specific
energy~$ \varepsilon $ is realized not for the zero angular momentum of
particles as in Schwarzschild case but for the values of the angular
momentum $l = - 2 A \varepsilon /(r_{\rm max}-2M)$.
    One obtains for given boundary value $\varepsilon = 1 / \sqrt{3} $ ($a=M$)
the limitation $ r < 3 M $.
    The limiting value $r= 3M$ takes place for $l=-2/\sqrt{3}$.
    For retrograde orbits with $\varepsilon = 5 / (3 \sqrt{3}) $ ($a=M$)
one has $ r < 27 M $.
    The limiting value $r= 27 M $ takes place for $l=-2/(15 \sqrt{3})$.

\vspace{1em}
%%%% *****************************************************************
{\centering  \section{The energy of collision of particles close to
the black hole}
\label{Sec4}}

    One can find the energy in the centre of mass frame of two colliding
particles $E_{\rm c.m.}$ with rest masses~$m_1$, $m_2$ taking the square of
    \begin{equation} \label{SCM}
\left( E_{\rm c.m.}, 0\,,0\,,0\, \right) = p^{\,i}_{(1)} + p^{\,i}_{(2)},
\end{equation}
    where $p^{\,i}_{(n)}$ are 4-momenta of particles $(n=1,2)$.
    Due to $p^{\,i}_{(n)} p_{(n)i}= m_n^2$ one has
    \begin{equation} \label{SCM2af}
E_{\rm c.m.}^{\,2} = m_1^2 + m_2^2 + 2 p^{\,i}_{(1)} p_{(2)i} .
\end{equation}
    Note that the energy of collisions of particles in the centre of mass
frame satisfies the condition
    \begin{equation} \label{Eb0}
E_{\rm c.m.} \ge m_1 + m_2.
\end{equation}
    This follows from the fact that the colliding particles move one
towards another with some velocities.

    It is important to note that $E_{\rm c.m.}$ for two colliding particles
is not a conserved value differently from energies of particles (relative
to infinity) $E_1$,  $E_2$.

    For the free falling particles with energies $E_1$,  $E_2$ and angular
momentum projections $J_1, J_2$ from~(\ref{geodKerr1})--(\ref{geodR})
one obtains~\cite{HaradaKimura11}:
    \begin{eqnarray}
E_{\rm c.m.}^{\,2} &\!=\!& m_1^2 + m_2^2 +
\frac{2}{\rho^2} \biggl[ \, \frac{P_1 P_2 -
\sigma_{1 r} \sqrt{R_1} \, \sigma_{2 r} \sqrt{R_2}}{\Delta}
\nonumber \\
&&-\, \frac{ (J_1 - a  E_1 \sin^2 \! \theta) (J_2 - a  E_2 \sin^2 \! \theta)}
{\sin^2 \! \theta}
\nonumber \\
&& -\, \sigma_{1 \theta} \sqrt{\Theta_1} \,
\sigma_{2 \theta} \sqrt{\Theta_2} \, \biggr].
\label{KerrL1L2}
\end{eqnarray}
    The big values of the collision energy can occur near the event horizon
if one of the particles has the ``critical'' angular moment $J_H$:
    \begin{eqnarray}
E_{\rm c.m.}^{\,2}(r \to r_H) = \frac{ (J_{1H} J_2 - J_{2H} J_1)^2}
{4 M^2 (J_{1H} - J_1) (J_{2H} - J_2)}
\nonumber \\
+\, m_1^2 \left[1+ \frac{J_{2H} \!- J_2}{J_{1H} \!- J_1}\right] +
m_2^2 \left[ 1+ \frac{J_{1H} \!- J_1}{J_{2H} \!- J_2}\right]
\label{GrPvPi}
\end{eqnarray}
(see Eq.~(16) in~\cite{GribPavlovPiattella2012}).
    This is the BSW effect.

    The energy in the centre of mass frame can be written
through the relative velocity~$ v_{\rm rel}$ of particles at the moment
of collision~\cite{GribPavlovPiattella2012}, \cite{Zaslavskii11}:
    \begin{equation} \label{Relsk03}
E_{\rm c.m.}^{\,2} = m_1^2 + m_2^2 +
\frac{2 m_1 m_2}{\sqrt{1 \!- v_{\rm rel}^2}}
\end{equation}
    and the nonlimited growth of the collision energy in the centre of mass
frame occurs due to growth of the relative velocity to the velocity of
light~\cite{Zaslavskii11}.

    The existence of particles moving from the gravitational radius in the
direction of larger $r$ along white hole geodesic can give us
the new opportunity for collisions independent of angular momentum
with non-limited energy near black holes.
    As we can see from~(\ref{KerrL1L2}) the difference between energy
of collisions with particle moving to increasing $r$  $ (\sigma_{2r} =1)$
and decreasing $r$  $ (\sigma_{2r} =-1)$ is
    \begin{equation} \label{DDD}
\Delta E_{c.m}^2 = \frac{4 \sqrt{R_1} \sqrt{R_2}} {\rho^2 \Delta} .
\end{equation}
    This difference is equal to zero for top point of trajectory,
when $R_2 =0$.
    But for non-critical particles $(J \ne J_H)$ the difference is infinite
large for collision on the horizon $(r \to r_H)$.
    So from~(\ref{KerrL1L2}) one has for such collisions
$(\sigma_{1r} \sigma_{2r} =-1)$ for $r\to r_H$: \,
$P_1 P_2 - \sigma_{1r} \sqrt{R_1} \sigma_{2r} \sqrt{R_2} > 0$,
\, $\Delta \to 0$,
    \begin{eqnarray} \label{ss12m}
E_{c.m} \sim \frac{ \sqrt{ 2 (P_1 P_2 - \sigma_{1r} \sqrt{R_1}
\sigma_{2r} \sqrt{R_2})}}{\rho \sqrt{\Delta}}
\nonumber \\
 \approx
\frac{a}{ \sqrt{r\!-\!r_H} } \sqrt{
\frac{2 (J_{1H} - J_1 ) (J_{2H} - J_2 ) }
{(r_H^2 \!+\! a^2 \cos^2 \theta )\sqrt{M^2\!-\!a^2} }} \to +\infty .
\end{eqnarray}
    The analogue of the result~(\ref{ss12m}) is valid not only for Kerr's black
holes but also for Schwarzschild black holes ($a=0, \sigma_1 \sigma_2 =-1$)
    \begin{equation} \label{SHw}
E_{c.m} \sim 2 \sqrt{ \frac{E_1 E_2}{ 1 - (r_H/ r)}} \to +\infty, \ \
r \to r_H=2M.
\end{equation}

\vspace{1em}
%%%% *****************************************************************
{\centering  \section{Conclusion}
\label{SecConcl}}

    {\bf a}) As it is clear black holes described by Schwarzschild and Kerr's
metric contrary to the widespread opinion due to the necessity of geodesics
completeness cannot be totally black.
    This means that in space external to the black hole there exist two
different types of geodesics with particles moving in them.
   One type which is often considered as the only existing type describes lines
going from the space-time infinity out of horizon inside this horizon with all
information lost in singularity.

    However close to the horizon there are other geodesics arriving from
the region inside it,   going from the horizon to external space but then
returning back inside the horizon.
    These lines originate in white hole singularity and so this singularity
cannot be ``naked''.
    This is some limitation of the cosmic censorship principle saying that
``naked'' singularities cannot be observed.
    In our paper we gave classification of all such lines making black hole
not black!

    {\bf b}) The collision of the particle moving along usual ``black hole''
geodesic with that moving on the ``white hole'' geodesic leads to the effect of
``supercollider'' so that the energy in the centre of mass frame can be very
large and new high energy physics of the Planckean scale can manifest itself.
    This is similar to the BSW effect considered in~\cite{BanadosSilkWest09}.
    As it was discussed by us previously~\cite{GribPavlov2007AGN}
hypothetical particles of dark matter can decay on light particles due to new
physics and ultra high energy particles can be observed on Earth in spite
of the high red shift on this way from the black hole
(similar consideration using Penrose effect see in~\cite{GarielSantosSilk14}).

    {\bf c}) In case of the physical collapse of the star usually it is
supposed that the geometry of the external to the horizon region is
described either by Schwarzschild or by Kerr's metric.
    So one expects existence of two types of geodesics in this case also.
    The difference is that absence of the white hole singularity in the
case of collapse leads to the problem of looking for some physical
mechanism of explanation of appearance of such lines in the process of the
collapse.
    This is an open problem to be solved in future.

\vspace{1mm}
%%%% *****************************************************************
{\bf Acknowledgments.}

    This work were supported by a grant from the John Templeton Foundation.

\vspace{11pt}
%%%%%%%%%%%%%%%%%%%%%%%%%%%%%%%%%%%%%%%%%%%%%%%%%%%%%%%%%%%%%%%%%%%%%%

\end{document}